\documentclass[twocolumn,showpacs,amssymb,aps,prc,nofootinbib,floatfix,superscriptaddress]{revtex4-1}
\usepackage{supertabular}
\usepackage{lipsum}
\usepackage{blindtext}
\usepackage{amsfonts}
\usepackage{tensor}
\usepackage{graphicx}
\usepackage{pict2e}
\usepackage{balance}
\usepackage{blindtext}

\usepackage{amssymb, amsmath,environ}

\usepackage{color}
\usepackage[colorlinks=true,urlcolor=black,linkcolor=black,citecolor=red]{hyperref}


\newcommand{\bse}{\begin{subequations}}
	\newcommand{\ese}{\end{subequations}}
\newcommand{\be}{\begin{equation}}
\newcommand{\ee}{\end{equation}}

\NewEnviron{eqsplit}{%
	\begin{equation}\begin{split}
	\BODY
	\end{split}\end{equation}
}

\makeatletter
\newcommand*\bigcdot{\mathpalette\bigcdot@{.5}}
\newcommand*\bigcdot@[2]{\mathbin{\vcenter{\hbox{\scalebox{#2}{$\m@th#1\bullet$}}}}}
\makeatother

\newcommand{\bea}{\begin{eqnarray}}
\newcommand{\eea}{\end{eqnarray}}
\newcommand{\ba}{\begin{array}}
	\newcommand{\ea}{\end{array}}

\newcommand{\nn}{{\nonumber}}

\def \beq{\begin{equation}}
\def \eeq{\end{equation}}
\def \beqa{\begin{eqnarray}}
\def \eeqa{\end{eqnarray}}

\begin{document}
%\preprint{MITP-23-014}		

\title{Rapidity-even directed flow splitting of protons and antiprotons as a probe of baryon stopping in relativistic heavy-ion collisions}

\author{Tribhuban Parida}
\email[]{tribhu.451@gmail.com}
\affiliation{Department of Physical Sciences,\\ Indian Institute of Science Education and Research Berhampur, Laudigam–760003, Dist.–Ganjam, Odisha, India}
\affiliation{AGH University of Krakow, Faculty of Physics and Applied Computer Science, aleja Mickiewicza 30, 30-059 Cracow, Poland}

\author{Sandeep Chatterjee}
\email[]{sandeep@iiserbpr.ac.in}
\affiliation{Department of Physical Sciences,\\ Indian Institute of Science Education and Research Berhampur, Laudigam–760003, Dist.–Ganjam, Odisha, India}

\begin{abstract}
We compare the rapidity-even directed flow $v_1^{\rm even}$ in Au+Au collisions at Beam Energy Scan (BES) energies for baryons and anti-baryons within a (3+1)-dimensional viscous relativistic hydrodynamics coupled to hadronic transport framework. The double-junction baryon stopping picture motivates a rapidity-even component in the baryon deposition in the initial state. We demonstrate that the split in the $v_1^{\rm even}$ of protons and anti-protons is sensitive to the rapidity extension of the baryon deposition that we associate with the double junction baryon stopping. Particularly, we find that the mid-rapidity curvature $\frac{d^2 \Delta v_1^{\rm even} (p-\bar{p})}{dy^2}\vert_{y=0}$ is a robust discriminator of the initial state baryon rapidity profiles. A simultaneous measurement of $\Delta v_1^{\rm even}$  and its curvature at mid-rapidity could  constrain both the baryon diffusion strength and the baryon
stopping profile, providing access to the physics of baryon stopping in relativistic heavy ion collisions.
\end{abstract}

\maketitle

\section{Introduction}
Exploring the QCD phase diagram and locating the QCD critical point are among the central objectives of current heavy-ion collision research~\cite{STAR:2013gus,STAR:2014egu,Stephanov:1998dy}. Achieving this requires reliable baseline predictions for experimental observables, against which any critical signal can be identified. Hydrodynamics has proven highly successful in describing a wide range of heavy-ion collision data~\cite{Busza:2018rrf,Schenke:2020mbo}, and is therefore a natural framework to extend to the finite baryon density regime. Such an extension not only provides the necessary baseline predictions but also offers a means to constrain the transport coefficients and the equation of state of QCD matter at finite baryon density~\cite{Denicol:2018wdp,Monnai:2026fkp,Monnai:2024pvy}.

However, one of the crucial inputs required for such hydrodynamic studies is a suitable initial condition, particularly the initial baryon density distribution. While initial conditions derived from transport models~\cite{JETSCAPE:2025wac,Du:2026rxk} are commonly employed, they fall short in efficiently reproducing all relevant observables. One such elusive observable is the rapidity dependent splitting of the directed flow $v_1$ between protons and antiprotons~\cite{STAR:2014clz}. Moreover, transport models offer limited physical insight into the mechanism of baryon stopping, as the complexity of these microscopic frameworks makes them difficult to interpret transparently.  Parametric initial conditions, on the other hand, are better suited in this regard, as they encode baryon stopping in an intuitive and physically transparent manner. A proper understanding of baryon stopping is therefore essential before constructing such initial conditions.

Recently, several models have been proposed that successfully capture this elusive $v_1$ splitting data~\cite{Parida:2025lhn,Du:2022yok,Bozek:2022svy}. Interestingly, models built on the baryon junction picture of hadrons~\cite{Du:2022yok} have demonstrated the ability to reproduce the experimental observations, lending support to the existence of the conjectured baryon junction picture. These models also probe the rapidity dependence of the stopping cross-sections, as well as the contributions from single and double  junction stopping mechanisms. It is therefore important to further investigate such models in order to constrain the initial baryon distribution, deepen our understanding of baryon stopping, and, most crucially, shed light on the role of baryon junctions in the stopping dynamics.

In this work, we study two parametrized baryon deposition profiles. In both profiles, a rapidity-even deposition component is associated with binary nucleon-nucleon collision profile, with the two profiles differing in the specific choice of the rapidity-even envelope function. This rapidity-even envelope can be naturally associated with the double junction stopping mechanism, since in a binary collision two units of baryon charge are involved.

The double baryon junction stopping cross-section was originally proposed to be rapidity-independent~\cite{Kharzeev:1996sq}, and this rapidity-independent (plateau) profile was adopted in earlier work to describe the proton $v_1$ data~\cite{Du:2022yok}. In our previous work~\cite{Parida:2022ppj,Parida:2022zse}, we relaxed this assumption by considering a rapidity-even but rapidity-dependent deposition profile, which still successfully reproduced the $v_1$ data for both protons and antiprotons. In the present work, our goal is to systematically compare these two classes of baryon deposition profiles: one corresponding to the plateau profile consistent with the rapidity-independent double junction stopping cross-section, and the other relaxing this assumption  through a rapidity-dependent envelope. A comparative study of experimental observables computed with these two distinct baryon profiles provides a phenomenological handle on the nature of the double baryon junction stopping cross-section and helps discriminate between the two pictures.

\section{Model}

The space-time evolution of the strongly interacting matter produced in
relativistic heavy-ion collisions is described within a hybrid framework
that couples viscous relativistic hydrodynamics to a hadronic transport
model. The hydrodynamic stage governs the evolution of the hot and dense
quark-gluon plasma phase, while the subsequent dilute hadronic stage, is handled by the transport model. The hydrodynamic evolution is seeded by parametrized initial profiles constructed using the Glauber model.

\subsection{Initial energy distribution}

The initial three-dimensional distribution of the energy density in the
thermalized fireball at a fixed proper time $\tau_0$ is given
by~\cite{Bozek:2010bi}
\beqa
  \epsilon(x,y,\eta_{s}; \tau_{0}) &=& \epsilon_{0} \left[ \left(
  N_{+}(x,y)\, f_{+}(\eta_{s}) + N_{-}(x,y)\, f_{-}(\eta_{s})
  \right)\right.\nn\\
  &&\left.\times \left( 1- \alpha \right) + N_{\rm bin}(x,y)\,
  \epsilon_{\eta_s}(\eta_{s})\, \alpha \right],
 \label{eq.tilt}
\eeqa
where $N_{+}(x, y)$ and $N_{-}(x, y)$ denote the surface densities of
forward- and backward-going participants at the transverse position
$(x, y)$, respectively, and $N_{\rm bin}(x, y)$ is the binary collision
density. The parameter $\alpha \in [0,1]$ controls the relative weight
of participant and binary collision contributions to the total deposited
energy. The rapidity envelope $\epsilon_{\eta_s}(\eta_s)$, which is even
in space-time rapidity $\eta_s$, is defined as
\begin{equation}
  \epsilon_{\eta_s}(\eta_s) = \exp \left(
  -\frac{\left(\vert\eta_{s}\vert - \eta_{0}\right)^2}{2\sigma_{\eta}^2}
  \;\theta\!\left(\vert\eta_{s}\vert - \eta_{0}\right) \right),
  \label{eq_etas_even_profile_for_epsilon}
\end{equation}
where $\eta_0$ and $\sigma_{\eta}$ are free parameters tuned to
reproduce the rapidity-differential charged-particle yield.

The asymmetric energy deposition structure in Eq.~\eqref{eq.tilt}
originates from the physical picture, introduced in
Ref.~\cite{Bozek:2010bi}, that a moving source preferentially deposits
energy along its direction of motion. This asymmetry is encoded in the
functions $f_{\pm}(\eta_s)$, defined as
\begin{equation}
    f_{\pm}(\eta_s) = \epsilon_{\eta_s}(\eta_s)\,\epsilon_{F,B}(\eta_s),
\end{equation}
with
\begin{equation}
    \epsilon_{F}(\eta_s) =
    \begin{cases}
    0, & \eta_{s} < -\eta_{m},\\[4pt]
    \dfrac{\eta_{s} + \eta_{m}}{2\eta_{m}}, & -\eta_{m} \le \eta_{s} \le \eta_{m},\\[6pt]
    1, & \eta_{s} > \eta_{m},
\end{cases}
\label{tilt_prof}
\end{equation}
and $\epsilon_{B}(\eta_s) = \epsilon_F(-\eta_s)$. This configuration
produces a source that is tilted within the reaction plane, and has been
shown to describe the rapidity-odd directed flow $v_1$ of charged
hadrons~\cite{Bozek:2010bi,Parida:2022lmt,Jiang:2021foj,Jiang:2021ajc}.

\subsection{Initial net-baryon distribution}
\label{baryon_sec}

The initial net-baryon density is parameterized
as~\cite{Parida:2022ppj}
\beqa
  n_{B}(x,y,\eta_s;\tau_{0}) &=& N_{B}\left[\left(
  N_{+}(x,y)\,f_{+}^{B}(\eta_{s})
  + N_{-}(x,y)\,f_{-}^{B}(\eta_{s})\right)\right.\nn\\
  &&\left.\times\left(1-\omega\right)
  + N_{\rm bin}(x,y)\,f^{B}_{\rm bin}(\eta_{s})\,\omega\right],
 \label{weight_ansatz_1_for_baryon}
\eeqa
where the rapidity envelope functions $f_{\pm}^{B}(\eta_s)$ describe the
longitudinal profile of net-baryon deposition from forward- and
backward-going
participants~\cite{Denicol:2018wdp,Shen:2020jwv}. Specifically,
\beqa
    f_{+}^{n_{B}}(\eta_s) &=&
    \theta\!\left(\eta_s - \eta_{0}^{n_{B}}\right)
    \exp\!\left[-\frac{\left(\eta_s - \eta_{0}^{n_{B}}\right)^2}
    {2\sigma_{B,+}^2}\right] \nn\\
    &&+\;\theta\!\left(\eta_{0}^{n_{B}} - \eta_s\right)
    \exp\!\left[-\frac{\left(\eta_s - \eta_{0}^{n_{B}}\right)^2}
    {2\sigma_{B,-}^2}\right]
\label{forward_baryon_envelop}
\eeqa
and
\beqa
    f_{-}^{n_{B}}(\eta_s) &=&
    \theta\!\left(\eta_s + \eta_{0}^{n_{B}}\right)
    \exp\!\left[-\frac{\left(\eta_s + \eta_{0}^{n_{B}}\right)^2}
    {2\sigma_{B,-}^2}\right] \nn\\
    &&+\;\theta\!\left(-\eta_s - \eta_{0}^{n_{B}}\right)
    \exp\!\left[-\frac{\left(\eta_s + \eta_{0}^{n_{B}}\right)^2}
    {2\sigma_{B,+}^2}\right].
\label{backward_baryon_envelop}
\eeqa
Here $\eta_{0}^{n_{B}}$ sets the peak rapidity of baryon deposition and
$\sigma_{B,\pm}$ control the asymmetric widths of the forward and
backward tails; both are tuned to the rapidity-differential net-proton
yield. The parameter $\omega \in [0,1]$ governs the relative contribution
of binary collisions to the baryon distribution~\cite{Parida:2022ppj}.
The binary component $f^{B}_{\rm bin}(\eta_s)$ is, by construction,
symmetric in $\eta_s$~\cite{Parida:2022ppj,Parida:2022zse}.

In this work we compare two distinct prescriptions for
$f^{B}_{\rm bin}(\eta_s)$, which will be referred to as the Gaussian
profile and the plateau profile, and examine their observable
consequences.

\begin{enumerate}

\item \textit{Gaussian profile}:
\begin{equation}
f^{B}_{\rm bin}(\eta_{s}) =
f_{+}^{n_{B}}(\eta_s) + f_{-}^{n_{B}}(\eta_s).
\end{equation}
This is the simplest rapidity-even envelope one can construct for the
binary component: it mirrors the prescription commonly adopted for
the energy deposition, where the $N_{\rm bin}$-weighted rapidity
envelope is taken as the sum of the forward and backward odd envelope
functions.

\item \textit{Plateau profile}:
\begin{equation}
f^{B}_{\rm bin}(\eta_{s}) =
\exp\!\left(-\frac{\left(\vert\eta_{s}\vert
- \eta_{0}^{n_B}\right)^2}{2\sigma_{B,+}^2}\;
\theta\!\left(\vert\eta_{s}\vert - \eta_{0}^{n_B}\right)\right).
\end{equation}
This form is motivated by the double-junction stopping
picture~\cite{Kharzeev:1996sq,Brandenburg:2022hrp}, in which baryon-stopping cross-section is approximately rapidity-independent
in the central region. The plateau-type profile was introduced in
this context in Ref.~\cite{Du:2022yok}, where it was shown to
capture the rapidity-odd $v_1$ splitting between protons and
antiprotons — an observable that is difficult to
reproduce. We note that in Ref.~\cite{Du:2022yok} this profile was
associated with a transverse distribution more peaked at the origin
than $N_{\rm part}$, rather than with $N_{\rm bin}$. In the present
framework, $N_{\rm bin}(x,y)$ plays an analogous role and we
accordingly associate the plateau profile with the binary collision
density.

\end{enumerate}

The primary goal of this comparison is to identify observables that are
sensitive to the choice of baryon deposition profile. Measurements that
can discriminate between the Gaussian and plateau forms would provide
direct phenomenological insight into the baryon stopping mechanism in
heavy-ion collisions and shed light on the conjectured baryon-junction
picture.

\subsection{Hydrodynamic evolution and transport coefficients}

The hydrodynamic evolution is initiated with the Bjorken flow ansatz
and carried out using the publicly available viscous relativistic
hydrodynamics code \textsc{music}~\cite{Schenke:2010nt,Schenke:2011bn,
Paquet:2015lta,Denicol:2018wdp}. Throughout the evolution, a constant
specific shear viscosity of $\eta/s = 0.08$ is assumed, while bulk
viscous corrections are neglected in the present study.

Baryon diffusion is incorporated through the baryon transport
coefficient $\kappa_B$, derived from the Boltzmann equation in the
relaxation time approximation~\cite{Denicol:2018wdp},
\begin{equation}
\kappa_{B} = \frac{C_B}{T}\, n_{B}
\left[ \frac{1}{3}\coth\!\left(\frac{\mu_B}{T}\right)
- \frac{n_B T}{\epsilon + \mathcal{P}} \right],
\end{equation}
where $n_B$ is the net-baryon density, $\epsilon$ and $\mathcal{P}$ are the local
energy density and pressure, $T$ is the temperature, and $\mu_B$ is the
baryon chemical potential. The parameter $C_B$ controls the overall
strength of baryon diffusion. The equation of state (EoS) employed here
imposes strangeness neutrality together with a fixed ratio of net-baryon
to net-charge density~\cite{Monnai:2019hkn}.

\subsection{Particlization and hadronic transport}

The conversion of the fluid into particles — particlization — is
performed on the hypersurface of constant energy density
$\epsilon_f = 0.26$~GeV/fm$^3$ using the
\textsc{iss} sampler~\cite{Shen:2014vra}. The phase-space distributions
of primordial hadrons on this hypersurface are evaluated following the
Cooper--Frye prescription~\cite{Cooper:1974mv}. The resulting hadrons are then passed to the hadronic transport code \textsc{urqmd}~\cite{Bass:1998ca,Bleicher:1999xi}, which simulates
elastic and inelastic scatterings in the dilute hadronic phase until
kinetic freeze-out is reached.

\subsection{Parameter selection}

The model parameters adopted in this work are largely inherited from our
previous study~\cite{Parida:2022zse}, where they were tuned to
simultaneously describe the rapidity-differential charged-particle yield,
the net-proton yield, and the rapidity-odd directed flow $v_1$ of
identified hadrons. The principal difference from that work is that the
present study employs event-by-event hydrodynamic simulations, which are
necessary to access the fluctuation-driven rapidity-even component of
directed flow. We have verified that the same parameter set used for previous 
study with event-averaged initial condition continues to
describe the experimental data well in this fluctuating setting.

Since the previous work was based on the Gaussian baryon deposition
profile, all parameters associated with the initial baryon distribution
are kept identical to Ref.~\cite{Parida:2022zse} when the Gaussian
profile is used. For the plateau profile, only the initial baryon
deposition parameters are readjusted to reproduce the same set of
calibration observables; the hydrodynamic transport coefficients and all
parameters governing the initial energy deposition are held fixed and
identical to the Gaussian case.

\section{Results}
\label{result_sec}

The rapidity-even directed flow $v_1^{\rm even}$ of charged hadrons has
been measured by the STAR Collaboration across a broad range of
collision energies in the Beam Energy Scan (BES) program~\cite{STAR:2018gji}. We compare
our model calculations against these measurements to validate the
framework before turning to the rapidity-even directed flow of
identified hadrons.

The rapidity-even $v_1$ is computed following the procedure of
Ref.~\cite{Luzum:2010fb}. In each event, the first-order event-plane angle
$\Psi_1$ is determined from the flow vector $Q$, defined as
\begin{equation}
Q\,e^{i\Psi_1} = \left\langle w_j\, e^{i\phi_j} \right\rangle,
\end{equation}
where the average $\langle\cdots\rangle$ runs over all reconstructed
tracks $j$ within the chosen kinematic acceptance, $\phi_j$ is the
azimuthal angle of track $j$, and the weight
\begin{equation}
w_j = (p_T)_j - \frac{\langle p_T^2 \rangle}{\langle p_T \rangle}.
\end{equation}

\begin{figure}
  \begin{center}
    \includegraphics[scale=0.35]{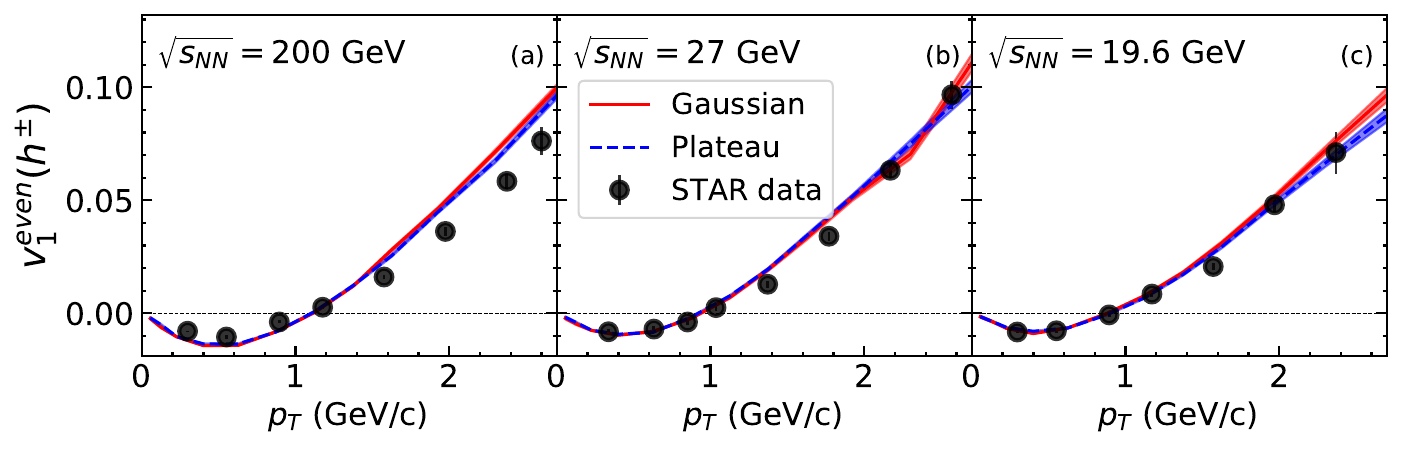}
    \caption{(Color online) Rapidity-even directed flow $v_1^{\rm even}$
    of charged hadrons as a function of $p_T$ for 0--10\% central Au+Au
    collisions at $\sqrt{s_{NN}} = 200$~GeV (A), 27~GeV (B), and
    19.6~GeV (C), for the two baryon deposition profiles with $C_B =
    1.0$. Model calculations (lines with bands) are compared with STAR
    measurements (filled circles)~\cite{STAR:2018gji}.}
    \label{fig1}
  \end{center}
\end{figure}

\begin{figure}
  \begin{center}
    \includegraphics[scale=0.35]{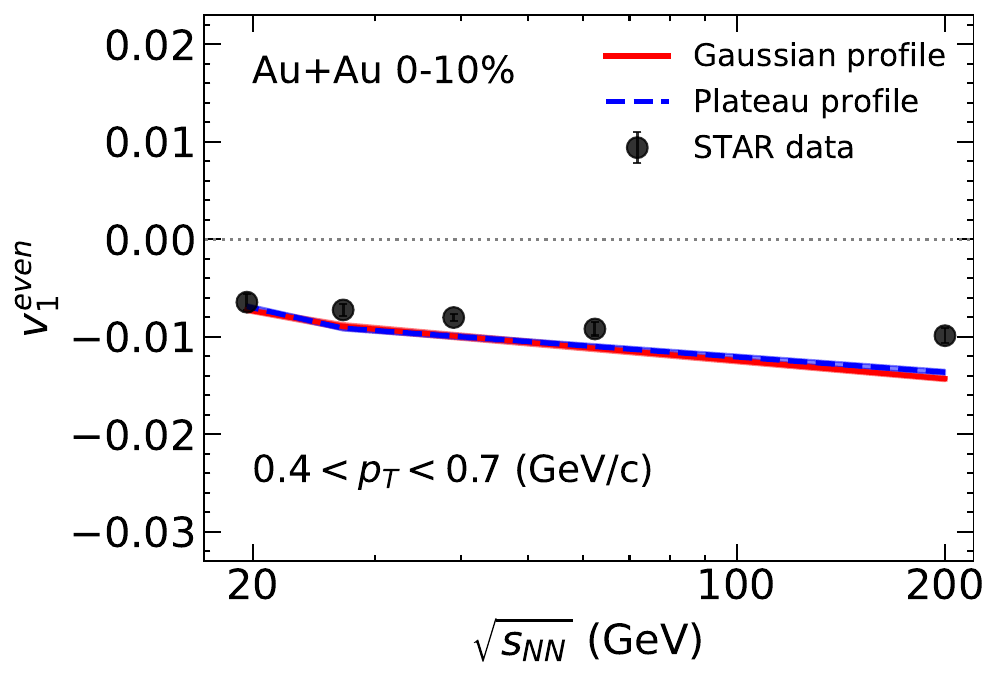}
    \caption{(Color online) $p_T$-integrated rapidity-even directed flow
    $v_1^{\rm even}$ of charged hadrons as a function of $\sqrt{s_{NN}}$
    for 0--10\% central Au+Au collisions. Model calculations (lines with
    bands) are compared with STAR measurements (filled circles)~\cite{STAR:2018gji}.}
    \label{fig2}
  \end{center}
\end{figure}

Figure~\ref{fig1} shows the $p_T$-differential $v_1^{\rm even}$ of
charged hadrons in 0--10\% central Au+Au collisions at
$\sqrt{s_{NN}} = 200$, 27, and 19.6~GeV, compared with STAR data~\cite{STAR:2018gji}. The
corresponding $p_T$-integrated $v_1^{\rm even}$ at mid-rapidity for the
same centrality class is shown as a function of collision energy in
Fig.~\ref{fig2}. In both cases, the same kinematic acceptance as in the
experimental analysis is applied to the model output. The calculations
describe the measured charged-hadron $v_1^{\rm even}$ well across the
full energy range considered, validating the framework and providing a
reliable baseline for the subsequent analysis of identified-hadron
rapidity-even directed flow.

The presence of finite baryon density gradients during the hydrodynamic
evolution induces a difference in the flow dynamics experienced by
baryons and antibaryons. This manifests as a splitting between baryon
and antibaryon observables — including yields, transverse momentum
spectra, and flow coefficients. Our framework already captures the
rapidity-odd $v_1$ splitting between protons and antiprotons; here we
extend this investigation to the rapidity-even sector.

\begin{figure}
  \begin{center}
    \includegraphics[scale=0.35]{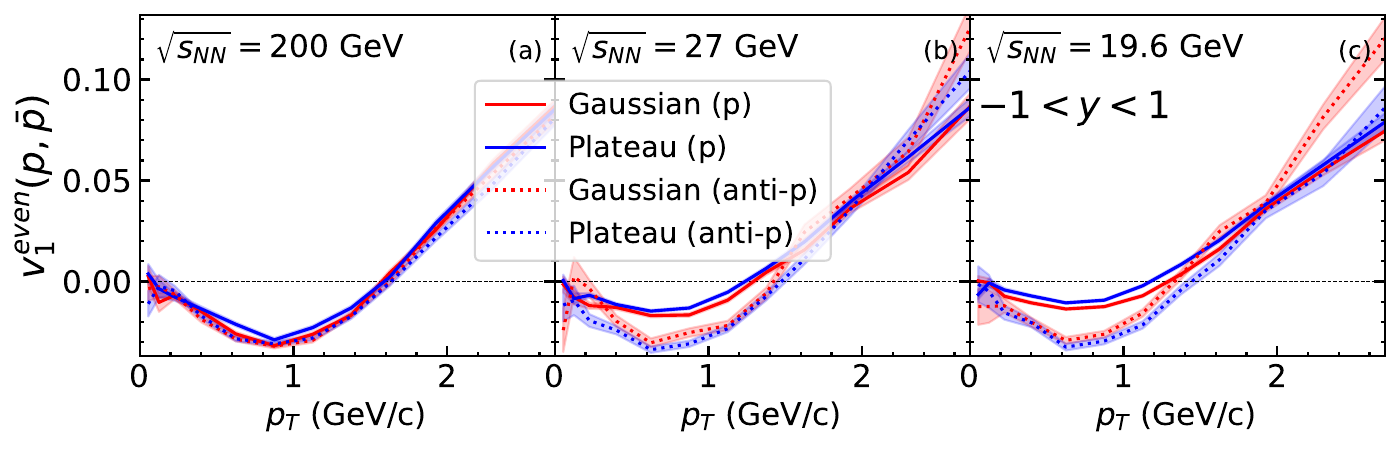}
    \caption{(Color online) $p_T$-differential rapidity-even directed
    flow $v_1^{\rm even}$ of protons (solid lines) and antiprotons
    (dotted lines) at mid-rapidity in 0--10\% central Au+Au collisions
    at $\sqrt{s_{NN}} = 27$~GeV. Results are shown for the Gaussian
    (red) and plateau (blue) baryon deposition profiles.}
    \label{fig3}
  \end{center}
\end{figure}

We focus on $\sqrt{s_{NN}} = 27$~GeV, where finite baryon density
effects are expected to be significant, and examine the
$p_T$-differential $v_1^{\rm even}$ of protons and antiprotons at
mid-rapidity. The results are shown in Fig.~\ref{fig3} for both the
Gaussian and plateau baryon deposition profiles. A clear splitting
between the proton and antiproton $v_1^{\rm even}$ is observed,
reflecting the baryon-antibaryon asymmetry generated by the diffusion
dynamics. Notably, however, the splitting is insensitive to the choice
of deposition profile: both the Gaussian and plateau prescriptions yield
nearly identical results in this observable.

\begin{figure}
  \begin{center}
    \includegraphics[scale=0.4]{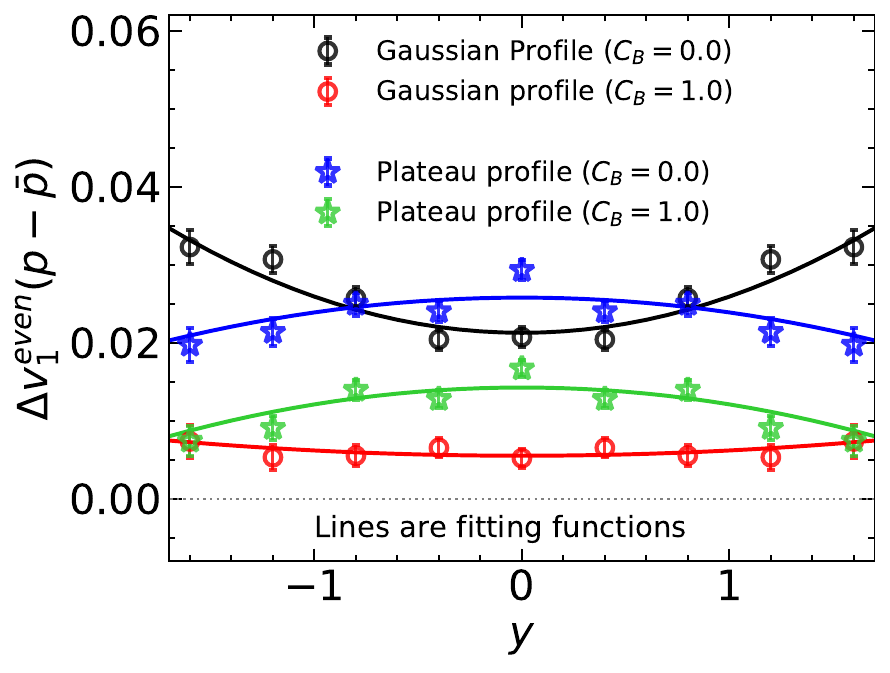}
    \caption{(Color online) The proton--antiproton splitting of the
    rapidity-even directed flow, $\Delta v_1^{\rm even}(y) = v_1^{\rm
    even}(p) - v_1^{\rm even}(\bar{p})$, as a function of rapidity in
    0--10\% central Au+Au collisions at $\sqrt{s_{NN}} = 27$~GeV.
    Results for the Gaussian profile are shown as circles (black:
    $C_B = 0$; red: $C_B = 1$) and for the plateau profile as stars
    (blue: $C_B = 0$; green: $C_B = 1$).}
    \label{fig_Deltav127}
  \end{center}
\end{figure}

Since the Gaussian and plateau profiles differ in their rapidity
structure of baryon deposition, one naturally expects a corresponding
difference in the rapidity dependence of the $v_1^{\rm even}$ splitting
between protons and antiprotons, $\Delta v_1^{\rm even}(y) \equiv
v_1^{\rm even}(p) - v_1^{\rm even}(\bar{p})$. This rapidity dependence
is shown in Fig.~\ref{fig_Deltav127} for 0--10\% central Au+Au
collisions at $\sqrt{s_{NN}} = 27$~GeV, for both profiles and for two
values of the baryon diffusion coefficient, $C_B = 0$ and $C_B = 1$.

The magnitude of $\Delta v_1^{\rm even}(y)$ varies with $C_B$, as
expected from the differing strength of baryon diffusion, but the
qualitative shape of the rapidity dependence remains the same within
each profile. The two profiles are, however, qualitatively distinct from
one another: the Gaussian profile produces a splitting with positive
curvature at mid-rapidity, whereas the plateau profile yields a
concave shape with negative curvature. This suggests that the
mid-rapidity curvature $\left.\frac{d^2\,\Delta v_1^{\rm even}}{dy^2}
\right|_{y=0}$ plays a role analogous to the mid-rapidity slope
$dv_1/dy$ in the rapidity-odd sector — a compact and experimentally
accessible quantity that encodes information about the underlying baryon
deposition mechanism.

\begin{figure}
  \begin{center}
    \includegraphics[scale=0.4]{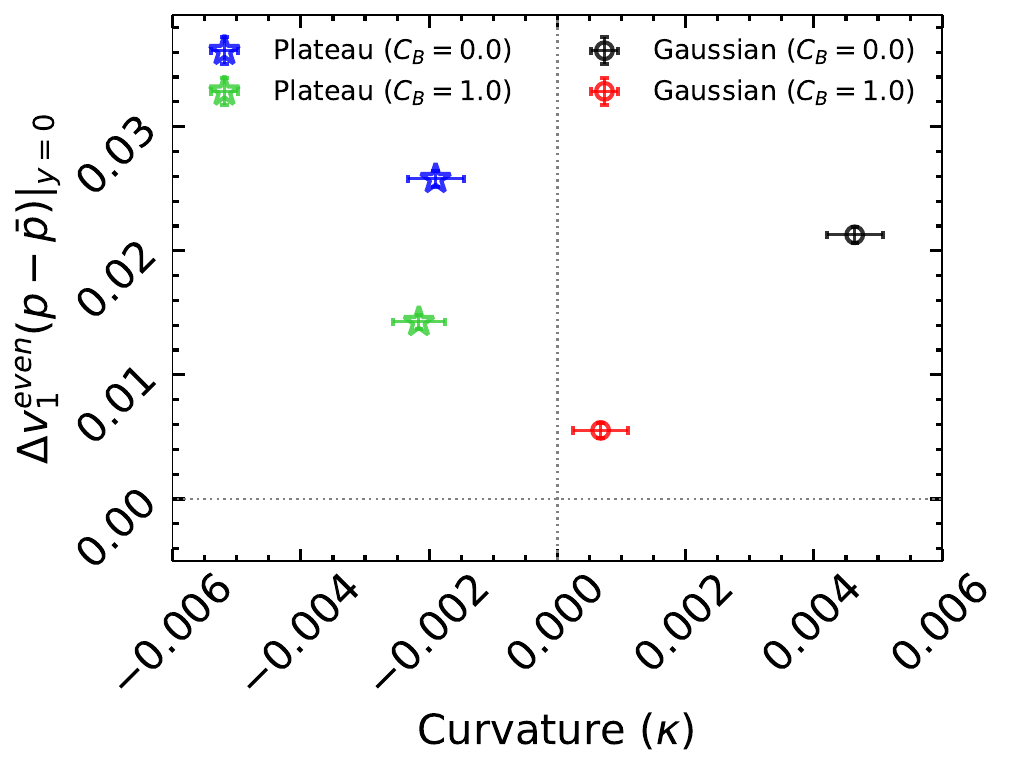}
    \caption{(Color online) Mid-rapidity value $a_0$ and curvature
    $\kappa$ extracted from a quadratic fit $a_0 + \kappa y^2$ to the
    rapidity-differential proton--antiproton $v_1^{\rm even}$ splitting
    $\Delta v_1^{\rm even}(y)$ of Fig.~\ref{fig_Deltav127}, for
    0--10\% central Au+Au collisions at $\sqrt{s_{NN}} = 27$~GeV.
    Results are shown for the Gaussian and plateau profiles and for
    different values of $C_B$.}
    \label{fig_curvature}
  \end{center}
\end{figure}

To present these two pieces of information — the mid-rapidity value and
the curvature — simultaneously and compactly, we parameterize
$\Delta v_1^{\rm even}(y)$ with a quadratic form $a_0 + \kappa y^2$ and
display the extracted coefficients in the two-dimensional plane
$({\kappa},\, a_0)$ in Fig.~\ref{fig_curvature}. The model points for
the Gaussian and plateau profiles, and for different values of $C_B$,
occupy clearly separated regions of this plane. A simultaneous
measurement of $a_0$ and $\kappa$ from experimental data would therefore
constrain both the baryon diffusion coefficient and the baryon deposition
profile, making this observable a particularly powerful probe of the
baryon stopping mechanism in heavy-ion collisions.

\section{Summary}

In this work, we have studied the rapidity-even directed flow
$v_1^{\rm even}$ of charged and identified hadrons in Au+Au collisions at BES energies within a hybrid framework coupling viscous relativistic
hydrodynamics (\textsc{music}) to hadronic transport (\textsc{urqmd}).
The initial conditions are constructed from the Glauber model,
with a tilted energy deposition profile, and a net-baryon
distribution that encodes the longitudinal stopping of baryons during
the collision.

A central element of this study is the comparison of two distinct
prescriptions of the initial baryon
deposition: a Gaussian profile, and a plateau profile, motivated by the hypothesis of double-junction baryon stopping, in which the baryon-stopping
cross-section is approximately rapidity-independent in the central
region. After validating the framework against STAR measurements of the
charged-hadron $v_1^{\rm even}$ across BES energies, we examined the
proton--antiproton splitting $\Delta v_1^{\rm even}(y) = v_1^{\rm
even}(p) - v_1^{\rm even}(\bar{p})$ and its rapidity dependence.

The key finding is that the two profiles produce qualitatively distinct
rapidity dependences of $\Delta v_1^{\rm even}(y)$: the Gaussian profile
yields a splitting with positive curvature at mid-rapidity, whereas the
plateau profile produces a concave shape with negative curvature. This
distinction is robust against variations in the baryon diffusion
strength $C_B$, meaning that the mid-rapidity curvature $\kappa =
\left.\frac{d^2\,\Delta v_1^{\rm even}}{dy^2}\right|_{y=0}$, together
with the mid-rapidity value $a_0$, can simultaneously constrain the
baryon diffusion coefficient and the baryon deposition profile. A
measurement of these quantities would thus provide direct phenomenological
insight into the baryon stopping mechanism and shed light on the
conjectured baryon-junction picture in heavy-ion collisions.

We note that the present study is phenomenological in nature and several
simplifying assumptions have been made. Pre-equilibrium dynamics prior
to the hydrodynamic initialization are not included, the initial
longitudinal flow profile is taken to be that of Bjorken flow. Incorporating these effects is expected to refine the quantitative predictions. Nevertheless, the primary goal of this work is to demonstrate the sensitivity of $\Delta v_1^{\rm even}(y)$ and its curvature to the underlying baryon deposition mechanism, and to establish this observable as a promising tool for constraining baryon stopping dynamics at BES energies. A more complete quantitative treatment is left for future work.

\section{Acknowledgement}
TP acknowledges support from the Polish Ministry of
Science and Higher Education and from the Polish National
Science Centre Grant No. 2023/51/B/ST2/01625.
	
\bibliography{cite.bib}

\end{document}